\documentclass[12pt]{article}

\usepackage{amssymb,amsmath,bm}
\renewcommand{\thefootnote}{\fnsymbol{footnote}}
\newcommand{\im}{{\rm Im\,}}

\newcommand{\tr}{{\rm Tr\,}}
\begin{document}
\title{}

\title{
\begin{flushright}
\begin{minipage}{0.2\linewidth}
\normalsize
WU-HEP-16-10 \\
EPHOU-16-006 \\*[50pt]
\end{minipage}
\end{flushright}
{\Large \bf 
Dynamical supersymmetry breaking on magnetized tori and orbifolds\\*[20pt] } }

\author{Hiroyuki~Abe$^{1,}$\footnote{
E-mail address: abe@waseda.jp}, \ 
Tatsuo~Kobayashi$^{2,}$\footnote{
E-mail address: kobayashi@particle.sci.hokudai.ac.jp} \ and \ 
Keigo~Sumita$^{1,}$\footnote{
E-mail address: k.sumita@aoni.waseda.jp
}\\*[20pt]
$^1${\it \normalsize 
Department of Physics, Waseda University, 
Tokyo 169-8555, Japan} \\
$^2${\it \normalsize 
Department of Physics, Hokkaido University, 
Sapporo 060-0810, Japan} \\*[50pt]}

\date{
\centerline{\small \bf Abstract}
\begin{minipage}{0.9\linewidth}
\medskip 
\medskip 
\small 
We construct several dynamical supersymmetry breaking (DSB) models 
within a single ten-dimensional supersymmetric Yang-Mills (SYM) theory, 
compactified on magnetized tori with or without orbifolding. 
We study the case that the supersymmetry breaking is triggered 
by a strong dynamics of $SU(N_C)$ SYM theory with $N_F$ flavors 
contained in the four-dimensional effective theory. 
We show several configurations of magnetic fluxes and orbifolds, 
those potentially yield, below the compactification scale, 
the field contents and couplings required for triggering DSB. 
We especially find a class of self-complete DSB models on orbifolds, 
where all the extra fields are eliminated by the orbifold projection 
and DSB successfully occurs within the given framework. 
Comments on some perspectives for associating the obtained DSB models
with the other sectors, such as the visible sector and another 
hidden sector for, e.g., stabilizing moduli, are also given. 
\end{minipage}}

\begin{titlepage}
\maketitle
\thispagestyle{empty}
\clearpage
\tableofcontents
\thispagestyle{empty}
\end{titlepage}

\renewcommand{\thefootnote}{\arabic{footnote}}
\setcounter{footnote}{0} 

\section{Introduction} 
Supersymmetric  models for particle physics have been quite 
actively studied for decades, 
and they will attract much more attention under the second season 
of Large Hadron Collider. 
The most famous and successful one is 
the minimal supersymmetric standard model (MSSM), which is 
indeed respected in many of supersymmetric models. 
In a generic model building, these supersymmetric models are accompanied by 
a sequestered hidden sector which breaks supersymmetry (SUSY) spontaneously, 
even when that is not mentioned explicitly. 
The SUSY breaking sector is certainly a key constituent of SUSY 
scenarios because of the fact that SUSY is broken in our real world 
at least below the electroweak scale. 

A wide variety of models for SUSY breaking sectors, solely or 
in association with the visible sector, have been proposed so far. 
In particular, many models of dynamical supersymmetry breaking (DSB) 
due to the strong dynamics of non-Abelian gauge theories, were proposed 
after the Seiberg duality revealed infrared behaviors of strongly coupled 
$\mathcal N=1$ SUSY theories~\cite{Seiberg:1994pq,Seiberg:1997vw}. 
These DSB scenarios are quite promising for completing SUSY models, 
because a large hierarchy between the Planck scale and the SUSY breaking 
scale (intrinsic strong scale) is easily generated by a logarithmic running of 
strong gauge couplings. 

In this paper, we construct DSB models in four-dimensional (4D) low-energy 
effective theory derived from ten-dimensional (10D) 
supersymmetric Yang-Mills (SYM) theories compactified on 
three 2-tori with magnetic fluxes. 
Extra-dimensional space with magnetic fluxes has been addressed 
as a hopeful candidate for the origin of flavor structure of the quarks and leptons, 
which is a big mystery of the standard model and its SUSY extensions. 
Magnetic fluxes on tori lead to gauge symmetry breaking 
and derive a product gauge group from a single large group, 
realizing generations of chiral fermions as degenerate zero-modes 
in the bi-fundamental representations~\cite{Bachas:1995ik,Cremades:2004wa} 
in the 4D effective theory. 

Indeed, a semi-realistic flavor structure was obtained in an MSSM-like model 
derived from the magnetized SYM theories~\cite{Abe:2012fj,Abe:2014soa}, 
where a suitable Yukawa hierarchy consistent with 
the observed masses and mixings of quarks and leptons is realized. 
This hierarchy is essentially due to the quasi-localization of 
wavefunctions in extra-dimensional space~\cite{ArkaniHamed:1999dc} 
caused by the magnetic fluxes. 
It was also shown that this model can be consistent with the recent 
experimental constraints on the Higgs boson mass and SUSY particle 
spectra, where a certain class of SUSY-breaking mediation mechanism 
is assumed~\cite{Abe:2014soa}. 

With $Z_2$ orbifolding~\cite{Abe:2008fi}, these attractive properties 
of magnetic fluxes remain still  and three-generation models 
of the quarks were studied on orbifolds~\cite{Abe:2008sx}. 
In Refs.~\cite{Abe:2014vza}, realistic Yukawa hierarchies 
were indeed realized on magnetized $Z_2 \times Z'_2$ orbifolds. 
These magnetized orbifolds\footnote{
Recently, $Z_3$, $Z_4$ and $Z_6$ orbifold models 
were also studied~\cite{Abe:2014noa,Abe:2015yva,Matsumoto:2016okl,Fujimoto:2016zjs}.} 
lead to a different flavor structure from the magnetized tori 
without orbifolding~\cite{Abe:2009vi,Abe:2009uz,Abe:2013bba}. 
Besides that, the orbifold projection can eliminate extra adjoint fields 
(those remain massless on tori and are phenomenologically disfavored 
in many cases), which would be a great advantage in a realistic model building. 

Thus, magnetized toroidal compactification with or without orbifolding 
is an exciting possibility of realizing the suitable visible (MSSM) sector 
in extra-dimensional field theories. 
As a second step towards completing these models, it is important to study 
SUSY breaking mechanisms on magnetized tori and orbifolds, which is the 
main purpose of this paper. 

The following sections are organized as follows. 

In Sec.~\ref{sec:2}, 
we review the 10D SYM theories compactified on magnetized tori. 
We adopt a  4D $\mathcal N=1$ description of 10D SYM theories, 
which is quite useful for the later model building. 
With this description, we give an overview of zero-mode configurations 
when the theory is compactified on three 2-tori with magnetic fluxes 
with/without $Z_2$ orbifolding. 

Sec.~\ref{sec:3} is the main part of this paper, 
where the construction of various DSB models is shown 
with several concrete magnetized backgrounds. 
In Sec.~\ref{sec:3.1}, we show certain aspects for DSB on magnetized tori 
with a simple configuration of magnetic fluxes which yields 
the gauge symmetry breaking $U(N) \rightarrow U(N_C) \times U(N_X)$, 
by assuming certain vacuum expectation values (VEVs) 
of the adjoint fields and their masses around them. 
First, a (metastable) DSB model is constructed in Sec.~\ref{sec:3.1.1} 
respecting the Intriligator-Seiberg-Shih (ISS) model~\cite{Intriligator:2006dd}, 
that is, $SU(N_C)$ SYM theory with $N_F$ fundamental massive quarks, 
satisfying $N_C-1 \le N_F<\frac32N_C$. 
We also construct a DSB model without massive quarks in Sec.~\ref{sec:3.1.2}, 
deriving a tadpole term from tri-linear couplings in the superpotential, 
via a suitable strong dynamics. 
In Sec.~\ref{sec:3.2}, we extend the flux configuration in such a way 
that the gauge symmetry breaking 
$U(N) \rightarrow U(N_C) \times U(N_X) \times U(N_Y)$ occurs. 
Then, we show a class of self-complete DSB models on magnetized orbifolds, 
where all the extra unwanted fields are eliminated by the orbifold 
projection and DSB successfully occurs within the given framework 
without any nontrivial assumptions. 
In Sec.~\ref{sec:3.3}, we comment on some perspectives for embedding the 
obtained DSB models into a single whole system including the visible 
(MSSM) sector and another hidden sector for the moduli stabilization. 

We conclude with the future prospects in Sec.~\ref{sec:4}. 

In Appendix~\ref{sec:app}, the other flux configurations are shown for 
deriving the same class of DSB models as the one demonstrated in Sec.~\ref{sec:3.2}.

\section{10D SYM theory on magnetized tori}
\label{sec:2}
We review 10D SYM theories on magnetized tori and orbifolds briefly, 
following Ref.~\cite{Abe:2012ya}. 
In this paper, the theories are compactified on $M^4\times T^2\times T^2\times T^2$ 
with/without $Z_2$ orbifolding. 
First, we introduce a 4D $\mathcal N=1$ description of higher dimensional SUSY 
theories which is quite useful for our model building. 
Using the description, 
we turn on Abelian magnetic fluxes in extra dimensional space 
and show an overview of zero-mode configurations on the magnetized tori. 
Finally, we explain about magnetized $Z_2$ orbifolds 
which are one of key ingredients in some of our DSB models. 

\subsection{4D $\mathcal N=1$ decomposition}
The 10D SYM theory consists of a 10D vector field $A_M$ ($M=0,1,\ldots,9$) 
and a 10D Majorana-Weyl spinor field $\lambda$. 
For the extra dimensional directions, 
we define complex coordinates~$z^i$ ($i=1,2,3$) 
and vectors~$A_i$ with complex structures~$\tau_i$ as 
\begin{equation*}
z^i\equiv \frac12\left(x^{2+2i}+\tau_ix^{3+2i}\right),\qquad 
A_i\equiv -\frac1{\im\tau_i}\left(\tau_i^*A_{2+2i}-A_{3+2i}\right). 
\end{equation*}
The periodic boundary conditions for the three 2-tori are given by 
$z_i\sim z_i+1$ and $z_i\sim z_i+\tau_i$. 
On this complex basis, the metric of three 2-tori is represented by 
\begin{eqnarray*}
ds_{\rm 6D}^2\equiv 2h_{\bar ij}d\bar z^{\bar i}dz^j,\qquad 
h_{\bar ij}=\delta_{\bar ij}2(2\pi R_i)^2, 
\end{eqnarray*}
where $R_i$ determines the period of $i$-th 2-torus. 

Now, the 10D vector field $A_M$ has been decomposed into 
a 4D vector and three complex scalar fields, $A_\mu$ and $A_i$. 
The spinor field can also be decomposed into four 4D Weyl spinors, 
which are distinguished by their chiralities on each 2-torus. 
We denote them as $\lambda_{+++}$, $\lambda_{+--}$, $\lambda_{-+-}$ 
and $\lambda_{--+}$ where the $i$-th subscript $\pm$ expresses 
the chirality on the $i$-th 2-torus, and the others 
(e.g., $\lambda_{---}$) are excluded by the 10D Weyl condition. 
We redefine these four spinors as 
\begin{equation*}
\lambda_0\equiv\lambda_{+++},\quad 
\lambda_1\equiv\lambda_{+--},\quad
\lambda_2\equiv\lambda_{-+-},\quad
\lambda_3\equiv\lambda_{--+}, 
\end{equation*}
for later convenience. 

These 4D component fields form 4D $\mathcal N=1$ supermultiplets, which 
are assigned to a vector~$V$ and three chiral superfields~$\phi_i$ as 
\begin{eqnarray}
V&\equiv&-\theta\sigma^\mu\bar\theta A_\mu+i\bar\theta\bar\theta\theta\lambda_0
-i\theta\theta\bar\theta\bar\lambda_0+\frac12\theta\theta\bar\theta\bar\theta D,\label{eq:vdef}\\
\phi_i&\equiv&\frac1{\sqrt2}A_i+\sqrt2\theta\lambda_i+\theta\theta F_i. \label{eq:phidef}
\end{eqnarray} 
The authors of Refs.~\cite{Marcus:1983wb,ArkaniHamed:2001tb} 
proposed an action in the 4D $\mathcal N=1$ superspace, 
that is equivalent to the usual component-action of 10D SYM theory 
with the definitions (\ref{eq:vdef}) and (\ref{eq:phidef}). 
Ref.~\cite{Abe:2012ya} showed its extension  to the toroidal 
compactifications where background magnetic fluxes are turned on. 
In the superspace formulation, a 4D $\mathcal N=1$ SUSY out of 
the full $\mathcal N=4$ SUSY possessed by 10D SYM theories 
becomes manifest, which is preserved by the configurations 
of magnetic fluxes. The $\mathcal N=1$ SUSY-preserving conditions 
are read from field equations for the auxiliary fields $D$ and $F_i$, 
those are shown later. 

\subsection{Zero-modes on magnetized tori}
Next we show the zero-mode structure on magnetized tori. 
In $U(N)$ gauge theory, magnetic fluxes on the $i$-th 2-torus can be 
represented by $N\times N$ matrix $M^{(i)}$ in 
\begin{equation*}
\langle A_i\rangle=\frac{\pi}{\im\tau_i}M^{(i)}\bar z_{\bar i}. 
\end{equation*}
We consider nonvanishing integer values for only diagonal entries 
of $M^{(i)}$, i.e., the Abelian magnetic fluxes. 
When some of them are degenerate, 
the gauge symmetry is broken as $U(N)\rightarrow U(N_a)\times U(N_b)\times\cdots$. 
We require these magnetic fluxes to satisfy conditions 
$\langle F_i\rangle=\langle D\rangle=0$ to preserve 4D $\mathcal N=1$ SUSY. 
These can be rewritten simply as~\cite{Abe:2012ya,Troost:1999xn}
\begin{equation}
\frac1{\mathcal A^{(1)}}M^{(1)}+\frac1{\mathcal A^{(2)}}M^{(2)}+
\frac1{\mathcal A^{(3)}}M^{(3)}=0, \label{eq:susy}
\end{equation}
where $\mathcal A^{(i)}$ represents the area of the $i$-th 2-torus. 
If this is not satisfied, SUSY is broken at a compactification scale 
which is, in general, extremely higher than the electroweak scale 
and some of SUSY particles get tachyonic masses due to 
$\langle D \rangle \ne 0$. 

In the following, we denote $(a,b)$-entries of $U(N)$ adjoint superfield 
$\phi_j$ by $\phi_j^{ab}$. For such bi-fundamental fields of 
$U(N_a) \times U(N_b)$, zero-mode equations on the magnetized 2-tori are given by 
\begin{eqnarray}
\left[\bar\partial_{\bar i} +\frac{\pi}{2\im\tau_i}
M_{ab}^{(i)}z_i \right]\phi_j^{ab} &=& 0 \qquad{\rm for}\quad i=j,
\label{eq:zeroii}\\
\left[\partial_i -\frac{\pi}{2\im\tau_i} M_{ab}^{(i)}\bar z_{\bar i}  \right]
\phi_j^{ab} &=& 0 \qquad{\rm for}\quad i\neq j, \label{eq:zeroij}
\end{eqnarray}
where $M_{ab}^{(i)} \equiv M_{a}^{(i)}-M_{b}^{(i)}$ expresses 
the difference between two diagonal entries in $M^{(i)}$. 
For positive values of $M_{ab}^{(i)}$, 
we find $M_{ab}^{(i)}$-degenerate zero-modes as 
solutions of Eq.~(\ref{eq:zeroii}), while Eq.~(\ref{eq:zeroij}) 
has no normalizable solution. 
On the other hand, for $M_{ab}^{(i)}<0$, only Eq.~(\ref{eq:zeroij}) 
allows $|M_{ab}^{(i)}|$-degenerate zero-modes. 
Thus, magnetic fluxes yield generations of chiral fermions.

\subsection{Magnetized orbifold}
We now consider $Z_2$ orbifolding on magnetized tori. 
The superfield description introduced above is compatible with 
orbifold projections, when we assign the same $Z_2$ parity to 
all the component fields contained in a single superfield. 
For example, we consider a $Z_2$ orbifold which acts on the 
first and the second 2-tori as $(z_1,z_2)\rightarrow (-z_1,-z_2)$. 
Under this $Z_2$ transformation, the superfields behave as 
\begin{eqnarray*}
V(x_\mu,z_1,z_2,z_3)&=&~~PV(x_\mu,-z_1,-z_2,z_3)P^{-1},\\
\phi_1(x_\mu,z_1,z_2,z_3)&=&-P\phi_1(x_\mu,-z_1,-z_2,z_3)P^{-1},\\
\phi_2(x_\mu,z_1,z_2,z_3)&=&-P\phi_2(x_\mu,-z_1,-z_2,z_3)P^{-1},\\
\phi_3(x_\mu,z_1,z_2,z_3)&=&~~P\phi_3(x_\mu,-z_1,-z_2,z_3)P^{-1},
\end{eqnarray*}
where the projection operator $P$ is given by an $N\times N$ matrix 
satisfying $P^2=1$. Then, all the elements are assigned into 
either even- or odd-parity mode under this $Z_2$ transformation. 

Orbifold projections reduce the number of degenerate 
zero-modes generated by magnetic fluxes, or eliminate them completely. 
Ref.~\cite{Abe:2008fi} identified the number of degeneracy of each 
$Z_2$-eigenmode with the sequence of magnetic fluxes on $Z_2$ orbifolds, 
that is shown in Table~\ref{tb:zeromode}. 
\begin{table}[th]
\center
\begin{tabular}{cccccccccccccc}
 $M$&$0$& $1$  &$2$  &$3$  &$4$  &$5$  &$6$  &$7$  &$8$  &$9$ & $10$&$2n$&$2n+1$ \\\hline
Even& $1$  &$1$  &$2$  &$2$  &$3$  &$3$  &$4$  &$4$  &$5$ & $5$  &$6$&$n+1$&$n+1$\\
Odd& $0$  &$0$  &$0$  &$1$  &$1$  &$2$  &$2$  &$3$  &$3$ & $4$  &$4$&$n-1$&$n$\\
\end{tabular}
\caption{This shows the number of active zero-modes on the magnetized orbifold.}
\label{tb:zeromode}
\end{table}

From these results, we find that the orbifold background gives variety 
to a magnetized model building. In the next section, we construct various 
DSB models on magnetized tori and orbifolds based on them.

\section{Dynamical supersymmetry breaking}
\label{sec:3}
In this section, we construct DSB models on a variety of magnetized background. 

First, we consider the simple configurations of magnetic fluxes 
leading to a gauge symmetry breaking $U(N)\rightarrow U(N_C)\times U(N_X)$, 
and show some specific configurations with which the resultant zero-modes 
contain certain DSB models such as the  ISS model~\cite{Intriligator:2006dd} 
and others. In the ISS model, for example, SUSY is broken by 
a strong dynamics of $SU(N_C)$ gauge theory with $N_F$ flavors. 
In our magnetized model building, $N_C$ and $N_F$ are determined by
 the degeneracies of Abelian magnetic fluxes and 
the degeneracies of the bi-fundamental zero-modes, respectively.
These models seem quite simple but there appear some extra massless modes, 
those should be eliminated or decoupled somehow to obtain successful DSB. 
As we will see, orbifold projections are not available for such a purpose, 
and we have to assume some extrinsic effects to eliminate the extra fields 
in this simple class of models. 

In the second part of this section, we consider more structural flux 
configurations that cause a gauge symmetry breaking 
$U(N) \rightarrow U(N_C)\times U(N_X)\times U(N_{X'})$. 
A great advantage of such extended configurations is that all the extra 
fields can be eliminated by a combination of magnetic fluxes and a certain 
orbifold projection, within a given framework of magnetized orbifold. 
They are really promising at least as long as we focus on the SUSY 
breaking sector. 

We finally discuss prospects of our DSB models in association with 
 other sectors, such as the visible (MSSM) and  other hidden 
(especially moduli stabilization) sectors.

\subsection{Models with $U(N) \rightarrow U(N_C)\times U(N_X)$}\label{sec:adj}
\label{sec:3.1}

\subsubsection{ISS-type models}
\label{sec:3.1.1}
In the first type of our model building, 
we try to realize  the ISS model~\cite{Intriligator:2006dd}, that is, 
the magnetized background is required to derive $SU(N_C)$ 
SYM theory with $N_F$ massive fundamental flavors from a 
single 10D $U(N)$ SYM theory. 
The IR description of such a model is given by 
\begin{equation*}
W= \lambda \phi_{in}\Phi^{ij}\bar\phi_j^n+\mu^2 \Phi^{ii}, 
\end{equation*}
where $\Phi$ and $\phi$ correspond to baryons and mesons 
($i,j=1,2,\ldots,N_F$ and $n=1,2,\ldots,N_C$). 
We can see that all the F-terms of $\Phi^{ij}$, 
$F_{\Phi_{ij}}\sim\mu^2\delta_{ij}+\lambda\phi_{in}\bar\phi_j^n$, 
cannot vanish simultaneously for $N_F>N_C$. 
This is the so-called rank-condition mechanism of SUSY breaking. 
In generic $SU(N_C)$ theories with $N_F$ flavors, 
SUSY breaking metastable vacua are realized within 
a range $N_C-1\leq N_F<\frac32N_C$. 
In particular, they can be long-lived 
when the quark mass scale is much smaller than 
the dynamical scale. 

We consider the configurations of magnetic fluxes which break 
the gauge symmetry as $U(N) \rightarrow U(N_C)\times U(N_X)$. 
For a while, we take both the $U(N_C)$ and $U(N_X)$ gauge groups 
to be non-Abelian ($N_C,N_X\geq2$) for the sake of generality. 
Such magnetic fluxes are given by 
\begin{eqnarray}
M^{(1)}&=&\begin{pmatrix}
0\times {\bm1}_{N_C}&0\\
0&M\times {\bm1}_{N_X}
\end{pmatrix},\quad 
M^{(2)}=\begin{pmatrix}
0\times {\bm1}_{N_C}&0\\
0&-1\times {\bm1}_{N_X}
\end{pmatrix},\nonumber\\
&&\hspace{6em}M^{(3)}=\begin{pmatrix}
0\times {\bm1}_{N_C}&0\\
0&0\times {\bm1}_{N_X}
\end{pmatrix},\label{eq:mag1}
\end{eqnarray}
where these matrices represent the $U(N_C+N_X)$ gauge space. 
This configuration preserves at least a 4D $\mathcal N=1$ SUSY 
with $\mathcal A^{(1)}/\mathcal A^{(2)}=M$ fixed 
for a positive value of $M$. 
The chirality projection caused by these magnetic fluxes eliminates the 
zero-modes of certain elements of the $U(N)$ adjoint chiral superfields, 
and we find for $M>0$, 
\begin{eqnarray*}
\phi_1&=&\begin{pmatrix}
\Xi_1&0\\
q&\Omega_1
\end{pmatrix},\quad 
\phi_2=\begin{pmatrix}
\Xi_2&\tilde q\\
0&\Omega_2
\end{pmatrix},\quad 
\phi_3=\begin{pmatrix}
\Xi_3&0\\
0&\Omega_3
\end{pmatrix}. 
\end{eqnarray*}
The magnetized background (\ref{eq:mag1}) induces $M$-pairs of vector-like 
quarks ($q,\tilde q$) in off-diagonal entries of $\phi_1$ and $\phi_2$. 
Diagonal entries, $\Omega_i$ and $\Xi_i$, 
correspond to $U(N_C)$ and $U(N_X)$ adjoint fields respectively. 

The 10D SYM theory allows couplings between $\phi_i$'s only in the form 
$\phi_1 \phi_2 \phi_3$ in the $\mathcal N=1$ superpotential. 
When the Wilson lines for $U(N_X)$ in the third 2-torus are somehow generated, 
they lead to a nonvanishing VEV of $\Omega_3$ and then the mass term 
$\langle\Omega_3\rangle q\tilde q$ is generated for the quarks.  
Then, the ISS-type DSB would occur if the quark mass scale is smaller 
than the dynamical scale of $SU(N_C)$ SYM theory. However, in order 
to realize the ISS model exactly, we need to further assume that the 
fluctuations of adjoint fields $\Omega_i$ and $\Xi_i$ around their VEVs 
($\langle\Omega_3\rangle \ne 0$, 
$\langle\Omega_{1,2}\rangle
=\langle \Xi_i \rangle =0$ in the present case) 
should be eliminated or decoupled from the DSB dynamics. 
Orbifold projections are not useful for such a purpose, because 
the nonvanishing (continuous) Wilson lines are generically incompatible 
with the orbifold background. Here, we just assume these extra fields 
obtain heavy masses due to some extrinsic effects from, 
e.g., supergravity/string corrections. 

Under the above assumption, we have two gauge theories with massive quarks: 
$SU(N_C)$ SYM with $M\times N_X$ fundamental flavors and 
$SU(N_X)$ SYM with $M\times N_C$ fundamental flavors. 
In the case with 
\begin{eqnarray}
N_C-1 &\leq& M \times N_X \ < \ \frac 32N_C, 
\label{eq:iss}
\end{eqnarray}
the ISS model is realized by the $SU(N_C)$ gauge theory. 
In this scenario, we have another constraint on the values of $N_C$ and $N_X$. 
The running of $SU(N_X)$ gauge coupling must be milder than that of $SU(N_C)$, 
which leads to the constraint 
\begin{eqnarray}
M\times N_X-3N_C &<& M\times N_C-3N_X 
\ \Leftrightarrow \ N_X < N_C. 
\label{eq:cond1}
\end{eqnarray}
While one can easily see that  both conditions~(\ref{eq:iss}) and 
(\ref{eq:cond1}) cannot be satisfied with $M=1$, 
it becomes easier for $M \geq 2$ to fulfill them 
and we can find many successful ansatzes, e.g., 
\begin{eqnarray}
N_C &=& 3,\qquad N_X \ = \ 2,\qquad M \ = \ 2. 
\nonumber
\end{eqnarray}

When the extra $U(N_X)$ gauge theory is Abelian, that is, $N_X=1$, 
we can realize  similar models much easier, because the second 
condition~(\ref{eq:cond1}) is not required in this case. Thus, 
we can construct DSB models concerning about only the first one~(\ref{eq:iss}).

\subsubsection{Models without massive quarks}
\label{sec:3.1.2}
We have another scenario on the magnetized background~(\ref{eq:mag1}) 
where the nonvanishing Wilson-lines are not required for DSB. 
As we have noticed, a key ingredient of this background is 
the following coupling, 
\begin{equation}
g\Omega_3q\tilde q =g\left(\langle\Omega_3\rangle +\tilde\Omega_3\right)
q\tilde q, 
\label{eq:omg3sp}
\end{equation}
where $g$ is a coupling constant. In the previous model, 
we have assumed a nonvanishing VEV $\langle\Omega_3\rangle$ 
and the absence of its fluctuation $\tilde\Omega_3$ at a low energy 
to realize the ISS-type DSB which has only the mass term for the vector-like 
quarks in the superpotential. 
Alternatively here we consider the case that 
$\tilde\Omega_3(=\Omega_3)$ is active while the Wilson-line 
$\langle\Omega_3\rangle$ is vanishing. 

Without the Wilson lines, the higher dimensional SYM theory 
does not produce any mass terms for $q$ and $\tilde q$ (as well as 
$\Omega_3$) at least at the leading order. This allows us to infer that, 
turning on a VEV $\langle\Omega_3\rangle \ne 0$ breaks some kinds of 
global symmetries of higher-dimensional SYM theory on magnetized tori, 
which prohibits the masses of bi-fundamental (as well as adjoint) fields. 
Thus, in the following, we can consider our models to be a chiral theory, 
as long as we do not introduce the continuous Wilson lines. 
This will become more clear in the next subsection. 

For the purpose to derive a DSB model without massive quarks, let us consider 
a situation where we can ignore the other block-diagonal entries of $\phi_i$ 
than $\Omega_3$ at a low energy (i.e., $\Omega_1$, $\Omega_2$ and $\Xi_i$ are 
decoupled) for simplicity. Again, this could not be realized by orbifolds 
because $\Omega_3$ and $\Xi_3$ have the same orbifold parity and both of them 
survive or vanish simultaneously under the orbifold projection. 
We should consider some extrinsic mechanisms to make the extra fields heavy 
as in the previous models. When they are somehow decoupled, the superpotential 
contains only the above Yukawa coupling~(\ref{eq:omg3sp}). 


In $SU(N_C)$ SYM theory with $N_f$ fundamental flavors, for $N_C>N_F$, 
the Affleck-Dine-Seiberg (ADS) potential~\cite{Davis:1983mz,Affleck:1983mk}
\begin{eqnarray}
W_{\rm ADS} &=& C_{N_C,N_F}\left(\frac{\Lambda^{3N_C-N_F}}
{{\rm det}\,\hat M}\right)^{1/(N_C-N_F)},
\nonumber
\end{eqnarray}
is obtained, where $\Lambda$ is the dynamical scale, 
$N_F\times N_F$ matrix $\hat M$ 
is defined as $\hat M^i_{~j}\equiv q^{in} \tilde q_{nj}$, 
and $C_{N_C,N_F}$ are constants. Our magnetized model contains 
$SU(N_C)$ SYM with $M\times N_X$ fundamental flavors and 
$SU(N_X)$ SYM with $M\times N_C$ fundamental flavors. 
We consider the case that the dynamics of the former 
non-Abelian gauge theory produces the above ADS potential, 
that is, $N_C>M\times N_X$. 
The total effective superpotential can be written 
in terms of the operator $\hat M$ as 
\begin{equation*}
W_{\rm effective}=g \tr\Omega_3\hat M
+C_{N_C,N_F}\left(\frac{\Lambda^{3N_C-N_F}}
{{\rm det}\,\hat M}\right)^{1/(N_C-N_F)}. 
\end{equation*}
This is almost the simplest DSB model found by 
Affleck, Dine and Seiberg~\cite{Affleck:1984xz}. 
This potential makes the operator $\hat M$ develop a nonvanishing VEV, 
$\langle \hat M\rangle\sim \Lambda$, 
and the resulting low-energy superpotential for $\Omega_3$ is 
\begin{eqnarray}
W=g\Lambda^2\Omega_3 +W_0, 
\nonumber
\end{eqnarray}
which is just like the Polonyi model~\cite{Polonyi:1977pj}. 

When the extra gauge theory is non-Abelian, $N_X\geq2$, 
we have to concern about the condition~(\ref{eq:cond1}) 
on $N_C$ and $N_X$, again. However, this is always satisfied 
when the ADS potential is generated , $N_C>M\times N_X$, 
for any positive value of $M$. As for Abelian cases $N_X=1$, 
such an extra constraint is not of course required. 
Thus, we can obtain a wide variety of this class of models 
as well as the previous ISS-type models discussed in Sec.~\ref{sec:3.1.1}. 

For $N_C=N_F$, the ADS potential is not generated. In this case, 
however, it is known that the strong dynamics induces chiral condensations 
yielding a vacuum with $\det\langle \hat M\rangle\neq0$. 
Therefore the Yukawa coupling~(\ref{eq:omg3sp}) produces a tadpole term 
for $\Omega_3$ breaking SUSY. 
Thus, a DSB model can be also obtained for $N_C= M\times N_X$. 
Here, the consistency condition~(\ref{eq:cond1}) requires $M\geq 2$.

\subsection{Models with $U(N) \rightarrow U(N_C) \times U(N_X) \times U(N_Y)$}
\label{sec:3.2}
So far, we have assumed that the extra adjoint fields are somehow decoupled. 
We here propose another class of DSB models on magnetized orbifold, where 
DSB successfully occurs within the given framework without requiring any 
extrinsic effects. We will find that more structural configurations of 
magnetic fluxes which cause a gauge symmetry breaking 
$U(N) \rightarrow U(N_C) \times U(N_X) \times U(N_Y)$ 
lead to self-complete DSB models on $Z_2 \times Z'_2$ orbifolds, 
where all the extra unwanted fields are eliminated 
below the compactification scale. 

We first explain an overview of this new class of models 
before giving a concrete configuration of magnetized background. 
Let us consider the gauge symmetry breaking due to 
magnetized backgrounds as $U(N)\rightarrow U(N_C)\times U(N_X)\times U(N_Y)$. 
Field contents responsible for the DSB dynamics here are ($i\neq j\neq k\neq i$) 
\begin{eqnarray}
\phi_i&=&\left(\begin{array}{c|c|c}
~~&~~&~~\\\hline
 & &S\\\hline
 & &\\
\end{array}\right),\quad 
\phi_j=\left(\begin{array}{c|c|c}
~~&\tilde Q&~~\\\hline
 & &\\\hline
 & &\\
\end{array}\right),\quad 
\phi_k=\left(\begin{array}{c|c|c}
~~&~~&~~\\\hline
 & &\\\hline
Q & &\\
\end{array}\right), \label{eq:fcon}
\end{eqnarray}
where three diagonal-blocks represent 
the product gauge group $U(N_C)\times U(N_X)\times U(N_Y)$ 
in $U(N)=U(N_C+N_X+N_Y)$ adjoint matrices and then 
the off-diagonal blocks in $\phi_i$'s are chiral multiplets 
in the corresponding bi-fundamental representations. 
Every mass term for $S$, $\tilde{Q}$ and $Q$ is forbidden 
by the (unbroken) gauge symmetry. 
In order to avoid chiral anomaly in adjunct $U(N_X)$ and $U(N_Y)$ 
gauge theories, we simply set $N_X=N_Y=1$ in the following. 
Even in this simple setup, the number of flavors can be controlled 
because the magnetic fluxes produce the degeneracy of zero-modes, 
enhancing the effective flavors. 

Chiral superfields $S$, $Q$ and $\tilde Q$ have a Yukawa coupling 
in the superpotential, 
\begin{eqnarray}
W &=& g SQ\tilde Q, 
\label{eq:sqq}
\end{eqnarray}
where $g$ expresses the effective coupling constant, which 
is given by an overlap integral of wavefunctions determined 
by magnetic fluxes and is calculable on magnetized tori 
(see~\cite{Cremades:2004wa,Abe:2008sx} for reviews). 
In accordance with the discussion in the previous subsection, 
for $N_C \geq N_F$, the $U(N_C)$ gauge dynamics enforces the operator 
$\hat M\equiv Q\tilde Q$ to develop a nonvanishing VEV, breaking SUSY.

\subsubsection{The essential structure}
We here aim to realize a minimal setup, that is, 
$N_F$ pairs of quarks ($Q$, $\tilde Q$) and a singlet $S$ in $U(N_C)$ SYM theory. 
We require the degeneracy of $S$ to be one in order to avoid the presence of 
extra massless fields. 

The generation structure of $Q$ and $\tilde Q$ should be produced 
on a single 2-torus. Otherwise, the rank of their Yukawa matrix is 
reduced and some fields become irrelevant to the DSB dynamics. 
Let us suppose that it is produced on the first 2-torus and denote 
magnetic fluxes felt by $Q$, $\tilde Q$ and $S$ by $M_1^Q$, 
$M_1^{\tilde Q}$ and $M_1^S$, respectively, where the subscript 
discriminates the first 2-torus. 
The gauge invariance enforces them to satisfy 
$M_1^Q+M_1^{\tilde Q}+M_1^S=0$. 
Furthermore, we find that only one of the three is positive and 
the others have to be negative. The reason is that the Yukawa 
coupling~(\ref{eq:sqq}) originates from the 10D gauge coupling 
$\phi_1\phi_2\phi_3$, and positive (negative) magnetic fluxes 
are required in order to produce zero-modes in $\phi_1$ ($\phi_{2,3}$) 
on the first 2-torus, which can be seen from Eqs.~(\ref{eq:zeroii}) 
and (\ref{eq:zeroij}). 

On a magnetized orbifold, $Q$, $\tilde Q$ and $S$ are assigned into 
either even- or odd-parity mode under the $Z_2$ transformation. 
The numbers of their zero-modes are determined by the magnetic fluxes 
$(M_1^Q,M_1^{\tilde Q},M_1^S)$ and their $Z_2$ parity. 
We show the relation between magnetic fluxes and the number of 
zero-modes on magnetized orbifolds in Table~\ref{tb:zeromode}. 
The $Z_2$ invariance of Yukawa coupling~(\ref{eq:sqq}) allows us 
to consider three cases for their $Z_2$ parity assignments, 
those are (even-even-even), (odd-odd-even) or (even-odd-odd) 
for $(Q,\tilde Q,S)$. 
Note that (odd-even-odd) is equivalent to (even-odd-odd) 
under the renaming $(Q,\tilde Q) \leftrightarrow (\tilde Q,Q)$, 
then we exclude the former. 

We eventually found only six patterns satisfy these conditions, 
which are shown in Table~\ref{tb:sixconf}. 
\begin{table}[h]
\center
\begin{tabular}{c|cc}
&$Z_2$ parity of $(Q,\tilde Q,S)$ &$(M_1^Q,M_1^{\tilde Q},M_1^S)$\\\hline
Pattern~1 & (even,\,even,\,even) & ($-n,\,n,\,0$) \\
Pattern~2 & (even,\,even,\,even) & ($-2n,\,2n+1,\,-1$) \\
Pattern~3 & (even,\,odd,\,odd) & ($-n,\,n+3,\,-3$) \\
Pattern~4 & (even,\,odd,\,odd) & ($-2n,\,2n+4,\,-4$) \\
Pattern~5 & (odd,\,odd,\,even) & ($-n,\,n,\,0$) \\
Pattern~6 & (odd,\,odd,\,even) & ($-2n-1,\,2n+2,\,-1$) \\
\end{tabular}
\caption{The six allowed combinations of 
$Z_2$ parity assignment and magnetic fluxes $(M_1^Q,M_1^{\tilde Q},M_1^S)$ 
are listed, where $n$ is an arbitrary positive integer. }
\label{tb:sixconf}
\end{table}

In order to produce the singlet $S$ without its multiplicity, 
$|M|=0,1$ unit of fluxes are allowed for the even-parity mode and 
$|M|=3,4$ for the odd-parity mode. The condition $M_1^Q+M_1^{\tilde Q}+M_1^S=0$ 
(one is positive and the other two are negative) severely restricts 
the values of $M_1^Q$ and $M_1^{\tilde Q}$, because the zero-mode number 
of $Q$ and that of $\tilde Q$ have to be equal for a successful DSB. 
Therefore, we can conclude that any other configurations are excluded. 

We find some differences among these six patterns. 
The first one is the value of coupling constants in $\lambda_{ij}SQ_i\tilde Q_j$. 
The Yukawa matrix $\lambda_{ij}$ is proportional to the identity matrix with 
Patterns~1 and 5. In the other cases, $\lambda_{ij}$ have nonvanishing values 
in their off-diagonal entries which can be calculated explicitly. 
The second difference is a constraint on the torus areas given by 
the SUSY preserving condition~(\ref{eq:susy}). 
In general, Patterns~2, 3, 4 and 6 yield characteristic values 
of the ratios $\mathcal A^{(1)}/\mathcal A^{(2)}$ and 
$\mathcal A^{(1)}/\mathcal A^{(3)}$ as one can see in Appendix~\ref{sec:app}. 
This might be of importance in combination with, especially, the visible sector. 
We will discuss about it in the last of this section. 

These six patterns allow us to construct several realistic DSB models 
within the given framework of magnetized orbifold without any nontrivial 
assumptions, which will be shown below.

\subsubsection{A self-complete model}
We propose concrete DSB models with explicit configurations of magnetic fluxes 
and orbifolds in the whole extra compact space. As discussed above, let us 
suppose that the main structure of our DSB models is produced on the first 2-torus. 
The configurations on the other two 2-tori are determined in order to eliminate all 
the extra field contents other than $Q$, $\tilde Q$ and $S$ in $\phi_1$, $\phi_2$ 
and $\phi_3$ without affecting the generation structure of them realized on the 
first 2-torus. The $Z_2$ parity assignments and the magnetic fluxes on the first 
2-torus are selected from the six patterns shown in Table~\ref{tb:sixconf} and 
the magnetic fluxes on the second and third 2-tori are enforced to satisfy the 
SUSY preserving condition~(\ref{eq:susy}). 

In the following, we construct an illustrating model on the basis of Pattern~1. 
That is, $Q$, $\tilde Q$ and $S$ are assigned into the even-parity mode 
on the first 2-torus with $(M_1^Q,M_1^{\tilde Q},M_1^S)=(-n,\,n,\,0)$. 
With the other five patterns, we can also realize similar models 
which we show in Appendix~\ref{sec:app}. 

Let us consider the following magnetized background, 
\begin{eqnarray}
M^{(1)}&=&\begin{pmatrix}
0&0&0\\
0&M&0\\
0&0&M
\end{pmatrix},\quad 
M^{(2)}=\begin{pmatrix}
0&0&0\\
0&-1&0\\
0&0&0
\end{pmatrix},\quad 
M^{(3)}=\begin{pmatrix}
0&0&0\\
0&0&0\\
0&0&-1
\end{pmatrix},\label{eq:realmag1} 
\end{eqnarray}
which breaks the $U(N)$ gauge symmetry down to 
$U(N_C) \times U(1)_X \times U(1)_Y$, 
while preserving $\mathcal N=1$ SUSY with 
$\mathcal A^{(1)}/\mathcal A^{(2)}=\mathcal A^{(1)}/\mathcal A^{(3)}=M$. 
We take the value of $M$ to be positive. 
In this case (before orbifolding) zero-mode contents are given by 
\begin{eqnarray*}
\phi_1&=&\begin{pmatrix}
\Xi_1&0& 0\\
\tilde Q'&\Xi'_1&0\\
Q&0&\Xi''_1
\end{pmatrix},\quad 
\phi_2=\begin{pmatrix}
\Xi_2&\tilde Q&0\\
0&\Xi'_2&0\\
0&S'&\Xi''_2
\end{pmatrix},\quad 
\phi_3=\begin{pmatrix}
\Xi_3&0&Q'\\
0&\Xi'_3&S\\
0&0&\Xi''_3
\end{pmatrix}.
\end{eqnarray*}

On this magnetized tori, we consider two $Z_2$ orbifold projections, 
i.e., a $Z_2 \times Z'_2$ orbifold. 
The first one, $Z_2$ orbifolding, acts on the first and the second 2-tori 
$(z_1,z_2,z_3)\rightarrow(-z_1,-z_2,z_3)$ with the projection operator
\begin{equation*}
P_{+--}=\begin{pmatrix}
+&0&0\\
0&-&0\\
0&0&-
\end{pmatrix}.
\end{equation*}
This operator successfully assigns the even-parity to all of 
$Q$, $\tilde Q$ and $S$ as in Pattern~1, while eliminating $S'$ 
and all the diagonal entries of $\phi_1$ and $\phi_2$. 
The second one is $Z'_2$ projection acting on the second and the third 2-tori 
$(z_1,z_2,z_3) \rightarrow (z_1,-z_2,-z_3)$ with the projection operator
\begin{equation*}
P_{+-+}=\begin{pmatrix}
+&0&0\\
0&-&0\\
0&0&+
\end{pmatrix}, 
\end{equation*}
which eliminates $Q'$, $\tilde Q'$ and all the diagonal entries of $\phi_3$. 
Consequently, the remaining zero-mode contents are exactly 
the ideal ones~(\ref{eq:fcon}). All the extra fields have been completely 
eliminated in the combination of magnetic fluxes and orbifolding. 

The total degeneracy of $S$ is certainly one. As for $Q$ and $\tilde Q$, 
their degeneracy is counted as the resulting number of $Z_2$ even modes 
with $|M|$ units of fluxes, which is read from Table~\ref{tb:zeromode}.  
We have obtained desirable $SU(N_C)$ SYM theory with $N_F=1,2,3,\ldots$ flavors, 
which can satisfy the condition $N_C=N_F\geq2$ or $N_C>N_F$ for a successful DSB. 
We show similar DSB models with Pattern~2 to 6 in Appendix~\ref{sec:app}. 

One might consider that the following configuration of magnetic fluxes 
is better than Eq.~(\ref{eq:realmag1}), 
\begin{eqnarray*}
M^{(1)}&=&\begin{pmatrix}
0&0&0\\
0&M&0\\
0&0&M
\end{pmatrix},\quad 
M^{(2)}=\begin{pmatrix}
0&0&0\\
0&-1&0\\
0&0&-1
\end{pmatrix},\quad 
M^{(3)}=\begin{pmatrix}
0&0&0\\
0&0&0\\
0&0&0
\end{pmatrix}. 
\end{eqnarray*}
The last 2-torus is vacant and the condition 
$\mathcal A^{(1)}/\mathcal A^{(2)}=M$ ensures that 
the SUSY is preserved. 
As for the gauge symmetry, $U(N)$ is broken down to $U(N_C)\times U(2)$. 
This $U(2)$ symmetry can be further broken to $U(1)_X \times U(1)_Y$ 
by orbifold projections. 
We again consider $Z_2 \times Z'_2$ orbifolds for this magnetized background. 
The first $Z_2$ acts on the first and the second 2-tori with the operator $P_{+--}$, 
and the second $Z'_2$ on the second and the third 2-tori with $P_{+-+}$. 
The surviving zero-modes are described as  
\begin{eqnarray}
\phi_1&=&\begin{pmatrix}
0&0&0\\
0&0&0\\
Q&0&0
\end{pmatrix},\quad 
\phi_2=\begin{pmatrix}
0&\tilde Q&0\\
0&0&0\\
0&0&0
\end{pmatrix},\quad 
\phi_3=\begin{pmatrix}
0&0&0\\
0&0&S\\
0&S'&0
\end{pmatrix}, \label{eq:fcon2}
\end{eqnarray}
and their full superpotential is found as 
\begin{equation}
W=gSQ\tilde Q, \label{eq:wpote}
\end{equation}
which has the same form as Eq.~(\ref{eq:sqq}). 

Although there is an extra massless field $S'$, it would not affect 
the DSB dynamics because this is a singlet under $U(N_C)$ and has 
no coupling in the superpotential.\footnote{We expect that this $S'$ 
can get a mass at loop-levels after DSB, because $Q$, $\tilde Q$ 
and $S$ have $U(1)_X$ gauge charges.} 
In this model, the zero-modes of both $S$ and $S'$ have no multiplicity, 
while the zero-mode degeneracies of $Q$ and $\tilde Q$ are equivalent to 
the previous model. 
Thus, the strong dynamics of $SU(N_C)$ gauge theory can generate DSB 
depending on the values of $N_C$ and $M$, without any nontrivial 
assumptions for extra field contents. 
This can be another self-sustained DSB model. 
Although this model contains a decoupled massless field $S'$, 
the model has a clear advantage to the previous one. 
There is no magnetic fluxes on the third 2-torus, then its area 
$\mathcal A^{(3)}$ is not constrained by SUSY conditions. 
This can be helpful for associating this model with the other sectors 
as we will discuss in the next subsection.

\subsection{Comments on the association with other sectors}
\label{sec:3.3}
We have constructed models for a SUSY breaking (hidden) sector, 
which must be combined with the MSSM (visible) sector and 
the other phenomenologically/cosmologically required sectors 
such as the moduli stabilization sector. 
Especially, when we consider a moduli stabilization mechanism 
based on non-perturbative effects such as gaugino condensations, 
like the Kachru-Kallosh-Linde-Trivedi (KKLT) scenario~\cite{Kachru:2003aw}, 
one or more strong gauge theories in the hidden sector are required 
for the non-perturbative dynamics. There are some key issues for 
combining these sectors altogether. 

Our models are based on SYM theories compactified on magnetized tori 
with/without orbifolds. The DSB models shown in subsection~\ref{sec:adj} 
is constructed without orbifolding, and thus, all the associated sectors 
such as the visible sector must also be constructed without orbifolding. 
On the other hand, the other DSB models on orbifolds have to be combined 
with the visible and the other sectors all constructed on the same orbifold. 
As the promising candidates for the visible sector, realistic flavor 
structures of MSSM-like models on magnetized tori~\cite{Abe:2012fj} and 
orbifolds~\cite{Abe:2008sx,Abe:2009vi} were derived so far. It is known 
that they are drastically different from each other. Therefore, we expect 
that models with or without orbifolding will be distinguishable phenomenologically. 

The values of higher-dimensional gauge coupling $g$, 
torus area $\mathcal A^{(i)}$ and complex structure $\tau_i$, are universal 
for all the sectors derived from the single higher-dimensional SYM theory. 
Thus, we have to choose common values for every sectors to be consistent 
with each other. We naively expect that most of these values are strongly 
constrained in the visible sector. 

First, the gauge coupling $g$ should be determined as follows. 
The 4D effective gauge coupling constant at the compactification scale, which 
is roughly given by a product of $g$ and the volume of extra compact space, 
must be consistent with the experimental data in the visible sector, 
i.e., the observed values of standard model (SM) gauge couplings. 
For example, if we consider MSSM for the visible sector, it automatically 
leads to a unified value of three SM gauge couplings at around $10^{16}$ GeV 
which is usually identified as the compactification scale, 
and the 4D effective gauge coupling can be fixed by the unified value. 
Next, the complex structures of tori are very important degrees of freedom to 
control the hierarchical structure of Yukawa couplings in the visible sector. 
Their values should be set to realize the quark and lepton masses 
and mixing angles~\cite{Abe:2012fj,Abe:2014vza}. 
Finally, the configurations of magnetic fluxes in the visible sector 
are extremely limited in order to realize the three generation structure of 
quarks and leptons, and the ratios of three torus areas, 
$\mathcal A^{(1)}/\mathcal A^{(2)}$ and/or $\mathcal A^{(1)}/\mathcal A^{(3)}$, 
are determined through the SUSY preserving conditions 
depending on the flux configuration. 

We remark that these constraints on parameters from the visible sector 
inevitably affect the model building for hidden sectors. 
Indeed, the DSB models shown in this paper also restrict the ratios 
$\mathcal A^{(1)}/\mathcal A^{(2)}$ and/or $\mathcal A^{(1)}/\mathcal A^{(3)}$, 
those must be consistent with the constraints from the visible sector. 
Therefore, the existence of unconstrained parameters in each sector is a 
great advantage, when we construct the whole system as a combination of 
the solely constructed visible and hidden sectors. Note that some of our 
DSB models with a vanishing flux in the third 2-torus restrict only one of 
the above two ratios. 

The models with two magnetized 2-tori (or even with a single magnetized 2-torus) 
among three are interesting from another point of view. We expect that our 
magnetized models based on 10D SYM theories would be completed being embedded 
into magnetized D9-brane systems, while the economically fluxed models have a 
potential to be compatible with D7-branes (or even D5-branes).\footnote{
It is argued that lower-dimensional D-branes may be derived from 
a magnetized D-brane in higher-dimensions with an infinite number 
of magnetic fluxes~\cite{Cremades:2004wa}. The effective field theory 
of such lower-dimensional branes can be derived based on 
such an argument~\cite{Abe:2015jqa}.}

When we construct the whole system by combining our DSB sector with the 
certain visible and other sectors, we also have to care about the direct 
couplings among them. All the sectors should be embedded into a single $U(N)$ 
gauge theory, if we regard our models as D-brane models with a single stack. 
On the other hand, with multi-stacks of D-branes (e.g., D3/D7 or D5/D9 
systems\footnote{The superfield formulation to describe such mixed D-brane 
systems was also constructed~\cite{Abe:2015jqa}.}), we can start from a product 
of multiple $U(N)$ gauge groups. 
In general, there exist bi-fundamental fields charged under two different sectors, 
depending on the configurations of magnetic fluxes and orbifolding. In particular, 
such bi-fundamental fields charged under the SM gauge groups are phenomenologically 
dangerous in many cases. We should also require that the strong dynamics of DSB 
and moduli stabilization sectors do not disturb each other through light fields 
charged under both sectors. 

Although these bi-fundamental fields are troublesome in generic cases, vector-like 
fields charged under both the MSSM and DSB sectors can be interesting, because 
they behave as messenger fields which mediate SUSY breaking contributions to the 
visible sector. In the previous analyses~\cite{Abe:2012fj} of magnetized models, 
it has been mostly assumed that the SUSY spectra are dominated by the 
moduli-mediation and/or anomaly-mediation~\cite{Randall:1998uk,Giudice:1998xp}, 
which depends on how to stabilize the moduli fields in association with 
the DSB sector. 
For example, in the KKLT-like moduli stabilization scenarios~\cite{Kachru:2003aw} 
with some concrete DSB sectors~\cite{Dudas:2006gr,Abe:2006xp}, contributions from 
the above two mediations can be comparable, and the so-called mirage mediation 
scenario~\cite{Choi:2004sx,Choi:2005ge,Endo:2005uy,Choi:2005uz} is realized. 
By assuming such a mediation scenario, the SUSY spectrum was studied 
in concrete magnetized models of the visible sector and some generic 
features were obtained~\cite{Abe:2012fj}. 
Then, it is interesting to employ one of our DSB models as the concrete 
hidden sector in this kind of scenario. The previous results can be 
deflected by the gauge-mediated contributions due to the appearance of 
messengers in the bi-fundamental representation between the MSSM and DSB sectors. 
We will study them in another places.

\section{Summary} 
\label{sec:4}
We have studied DSB models within the framework of 10D SYM theories 
compactified on magnetized tori and orbifolds. 

First, aspects for DSB on magnetized tori/orbifolds have been shown 
with the simple configurations of magnetic fluxes which causes 
the gauge symmetry breaking $U(N) \rightarrow U(N_C)\times U(N_X)$, 
by assuming (non)vanishing VEVs of adjoint fields and (non)decoupling 
of their fluctuations from the DSB dynamics around the VEVs. 
Then, in order for the strong dynamics of $SU(N_C)$ SYM theories with $N_F$ 
flavors to trigger a dynamical SUSY breaking, certain relations between 
$N_C$ and $N_F$ are required. It is remarkable that the number of flavors 
$N_F$ can be controlled by magnetic fluxes in our model, in other words, 
the background flux configuration determines whether DSB occurs or not. 

At the same time, however, we also find that the decoupling of some extra 
adjoint fields, those could not be eliminated by orbifold projections 
in the model building procedures, 
is necessarily assumed in this simple class of models. Otherwise the 
existence of them could spoil the successful DSB and/or are already 
ruled out by phenomenological/cosmological observations. 
In the case that some extrinsic mechanisms realize the assumed situations, 
these DSB models on magnetized tori are available for a further model building, 
while the decoupling of extra adjoint fields is in general a challenging issue 
in the model building based on SYM theories in higher-dimensional spacetime. 

Then, next, we have proposed another class of DSB models on orbifolds by 
extending the previous configurations of magnetic fluxes which preserve 
$U(N_C) \times U(N_X)$ symmetry to those yield 
$U(N_C) \times U(N_X) \times U(N_Y)$, 
especially, to the simplest one $U(N_C) \times U(1)_X \times U(1)_Y$. 
We have searched such flux configurations that the $SU(N_C)$ SYM theory 
contains $N_F$ vector-like pairs $(Q,\,\tilde Q)$ with their 
nonvanishing Yukawa couplings to a singlet $S$. 
As the result, we found six patterns of suitable configurations. 

On the basis of one of these six patterns, we demonstrated the construction of 
a self-complete DSB model on a $Z_2 \times Z'_2$ orbifold, where all the extra 
fields below the compactification scale are eliminated by the combination of 
chiral projections due to magnetic fluxes and the orbifold projections. 
In Appendix~\ref{sec:app}, we also show the other five patterns allow us to 
construct similar feasible models. Therefore, we conclude that, in this 
class of magnetized orbifold models, we can realize DSB without relying on 
any extrinsic mechanisms to eliminate extra fields. 

Furthermore, we have studied another choice of magnetic fluxes, where 
only two of the three 2-tori are fluxed. Although this permits a presence 
of one more singlet $S'$ without the Yukawa couplings to quarks, 
it of course does not disturb the DSB dynamics and can be another 
self-sustained DSB model. It is remarkable that the existence of unfluxed 
2-torus can be an advantage when we combine the DSB (hidden) sector with 
the MSSM (visible) sector~\cite{Abe:2015jqa}. 

As discussed in the previous section, our DSB models should be embedded into 
a larger unified system being compatible with the MSSM sector and the others, 
e.g., the moduli stabilization sector. This must be an important task from 
both theoretical and phenomenological points of view. We expect some of 
the six patterns we found and their extensions being suitable for such embeddings. 
Finally, it is an interesting possibility that such a whole system is realized 
by magnetized D-branes. In this case, we should verify some stringy consistency 
of the full system containing all the sectors for completing our scenario on 
magnetized tori/orbifolds. These are remained as future works.

\subsection*{Acknowledgements}
The authors would like to thank Yuji Omura and Tokihiro Watanabe 
for valuable discussions which initiate this work. 
H.~A. was supported in part by the Grant-in-Aid for Scientific 
Research No.~25800158 from the Ministry of Education, Culture, 
Sports, Science and Technology (MEXT) in Japan. 
T.~K. was supported in part by the Grant-in-Aid for Scientific 
Research No.~25400252 from the MEXT in Japan.

\appendix
\section{Other self-complete models with Pattern~2 to 6}
\label{sec:app}
We have shown a concrete DSB model based on Pattern~1 shown 
in Table~\ref{tb:sixconf}. We find that similar models can be 
realized with the other patterns and demonstrate them here 
(Models shown here contain just $N_F$-pairs of $(Q,\,\tilde Q)$ 
and one singlet $S$.). 

We start from Pattern~2, where all of $Q$, $\tilde Q$ and $S$ 
are assigned into the even-parity mode, and their fluxes are 
parametrized as $(M_1^Q,M_1^{\tilde Q},M_1^S)=(-2n,\,2n+1,\,-1)$ 
with a positive integer $n$. 
The suitable magnetic fluxes are given by 
\begin{eqnarray*}
M^{(1)}&=&\begin{pmatrix}
0&0&0\\
0&2n&0\\
0&0&2n+1
\end{pmatrix},\quad 
M^{(2)}=\begin{pmatrix}
0&0&0\\
0&-1&0\\
0&0&0
\end{pmatrix},\quad 
M^{(3)}=\begin{pmatrix}
0&0&0\\
0&0&0\\
0&0&-1
\end{pmatrix}, 
\end{eqnarray*}
which break gauge symmetry as 
$U(N)\rightarrow U(N_C)\times U(1)\times U(1)$ 
and satisfy the SUSY preserving condition with 
$\mathcal A^{(1)}/\mathcal A^{(2)}=2n$ and 
$\mathcal A^{(1)}/\mathcal A^{(3)}=2n+1$. 
We need two different orbifold projections to 
eliminate extra fields completely, then consider 
a $Z_2 \times Z'_2$ orbifold. 
The $Z_2$ orbifolding acts on the first and the second tori 
with the projection operator $P_{+--}$, while the $Z_2'$ 
orbifolding acts on the second and the third tori with 
the projection operator $P_{+-+}$. 
These are consistent with the parity assignment of Pattern~2 
and eliminate all the extra entries of $\phi_i$. 
The net number of zero-mode of $S$ is one. 
That of $Q$ ($\tilde Q$) is identified as the number 
of even-parity mode with $|M|=2n$ ($2n+1$) fluxes. 
We see from Table~\ref{tb:zeromode} that both the degeneracies 
of $Q$ and $\tilde Q$ are equal to $n+1$. 

With Pattern~3, $(Q,\tilde Q,S)$ are assigned into the 
(even,odd,odd)-parity mode, and their fluxes are given by 
$(-n,\,n+3,\,-3)$ with a positive integer $n$. 
A similar model is obtained on the same $Z_2 \times Z'_2$ 
orbifold as Pattern~2 but with the different magnetic fluxes, 
\begin{eqnarray*}
M^{(1)}&=&\begin{pmatrix}
0&0&0\\
0&n&0\\
0&0&n+3
\end{pmatrix},\quad 
M^{(2)}=\begin{pmatrix}
0&0&0\\
0&-1&0\\
0&0&0
\end{pmatrix},\quad 
M^{(3)}=\begin{pmatrix}
0&0&0\\
0&0&0\\
0&0&-1
\end{pmatrix}. 
\end{eqnarray*}
The net number of $Q$ ($\tilde Q$) is identified as that of 
even-parity (odd-parity) mode with $n$ ($n+3$) fluxes. 
We find the degeneracies of $Q$ and $\tilde Q$ are equal to each other. 

With Pattern~4, $(Q,\tilde Q,S)$ are assigned into the 
(even,odd,odd)-parity mode, and their fluxes are given by 
$(-2n,\,2n+4,\,-4)$ with a positive integer $n$. 
A similar model is obtained with the following magnetic fluxes, 
\begin{eqnarray*}
M^{(1)}&=&\begin{pmatrix}
0&0&0\\
0&2n&0\\
0&0&2n+4
\end{pmatrix},\quad 
M^{(2)}=\begin{pmatrix}
0&0&0\\
0&-1&0\\
0&0&0
\end{pmatrix},\quad 
M^{(3)}=\begin{pmatrix}
0&0&0\\
0&0&0\\
0&0&-1
\end{pmatrix}, 
\end{eqnarray*}
on the $Z_2 \times Z'_2$ orbifold, 
where $Z_2$ acts on the first and the second tori with $P_{+-+}$, 
and $Z_2'$ acts on the second and the third tori with $P_{+-+}$. 
The net number of $Q$ ($\tilde Q$) is identified as that of 
even-parity (odd-parity) mode with $2n$ ($2n+4$) fluxes. 
We find both the degeneracies of $Q$ and $\tilde Q$ are $n+1$. 

With Pattern~5, $(Q,\tilde Q,S)$ are assigned into the 
(odd,odd,even)-parity mode, and their fluxes are given by 
$(-n,\,n,\,0)$ with a positive integer $n$. 
A similar model is obtained with the following magnetic fluxes, 
\begin{eqnarray*}
M^{(1)}&=&\begin{pmatrix}
0&0&0\\
0&n&0\\
0&0&n
\end{pmatrix},\quad 
M^{(2)}=\begin{pmatrix}
0&0&0\\
0&-1&0\\
0&0&0
\end{pmatrix},\quad 
M^{(3)}=\begin{pmatrix}
0&0&0\\
0&0&0\\
0&0&-1
\end{pmatrix}, 
\end{eqnarray*}
on the $Z_2 \times Z'_2$ orbifold, 
where $Z_2$ acts on the first and the second tori with $P_{+++}$, 
and $Z_2'$ acts on the second and the third tori with $P_{+-+}$. 
Both the net numbers of $Q$ and $\tilde Q$ are equal to that of 
odd-parity mode with $n$ fluxes. 

Finally with Pattern~6, $(Q,\tilde Q,S)$ are assigned into the 
(odd,odd,even)-parity mode, and their fluxes are parametrized as 
$(-2n-1,\,2n+2,\,-1)$ with a positive integer $n$. 
A similar model is obtained on the same $Z_2 \times Z'_2$ 
orbifold as Pattern~5 but with the different magnetic fluxes, 
\begin{eqnarray*}
M^{(1)}&=&\begin{pmatrix}
0&0&0\\
0&2n+1&0\\
0&0&2n+2
\end{pmatrix},\quad 
M^{(2)}=\begin{pmatrix}
0&0&0\\
0&-1&0\\
0&0&0
\end{pmatrix},\quad 
M^{(3)}=\begin{pmatrix}
0&0&0\\
0&0&0\\
0&0&-1
\end{pmatrix}. 
\end{eqnarray*}
The net number of $Q$ ($\tilde Q$) are identified as that of 
odd-parity mode with $|M|=2n+1$ ($2n+2$) fluxes. 
Both the degeneracies of $Q$ and $\tilde Q$ are equal to $n$.

\end{document}